\documentclass[preprint2]{emulateapj}
\usepackage{natbib}
\usepackage{mathrsfs}
\usepackage{threeparttable}

\def\ha{H\footnotesize${\alpha}$\normalsize}
\def\lya{Ly$\alpha$ }
\def\ergcms{erg~cm$^{-2}$~s$^{-1}$}
\def\ergs{erg~s$^{-1}$}



\bibpunct[;]{(}{)}{,}{a}{}{;}

\shorttitle{The Radiated Energy Budget of Chromospheric Plasma in a Major Solar Flare Deduced From Multi-Wavelength Observations}

\shortauthors{Milligan et al.}

\begin{document}

\title{The Radiated Energy Budget of Chromospheric Plasma in a Major Solar Flare Deduced From Multi-Wavelength Observations}
\notetoeditor{the contact email is r.milligan@qub.ac.uk, is the only one which should appear on the journal version}
\author{Ryan O. Milligan\altaffilmark{1,2,3}, Graham S. Kerr\altaffilmark{4}, Brian R. Dennis\altaffilmark{2}, Hugh S. Hudson\altaffilmark{4,5}, Lyndsay Fletcher\altaffilmark{4}, Joel C. Allred\altaffilmark{2}, Phillip C. Chamberlin\altaffilmark{2}, Jack Ireland\altaffilmark{2,6}, Mihalis Mathioudakis\altaffilmark{1}, \& Francis P. Keenan\altaffilmark{1}}

\altaffiltext{1}{Astrophysics Research Centre, School of Mathematics \& Physics, Queen's University Belfast, University Road, Belfast, Northern Ireland, BT7 1NN}
\altaffiltext{2}{Solar Physics Laboratory (Code 671), Heliophysics Science Division, NASA Goddard Space Flight Center, Greenbelt, MD 20771, USA}
\altaffiltext{3}{Department of Physics, Catholic University of America, 620 Michigan Avenue, Northeast, Washington, DC 20064, USA}
\altaffiltext{4}{School of Physics and Astronomy, SUPA, University of Glasgow, Glasgow G12 8QQ, UK}
\altaffiltext{5}{Space Sciences Laboratory, UC Berkeley, 7 Gauss Way, Berkeley CA USA 94720-7450}
\altaffiltext{6}{ADNET Systems, Inc., Solar Physics Laboratory (Code 671), Heliophysics Science Division, NASA Goddard Space Flight Center, Greenbelt, MD, 20771, USA}

\begin{abstract}
This paper presents measurements of the energy radiated by the lower solar atmosphere, at optical, UV, and EUV wavelengths, during an X-class solar flare (SOL2011-02-15T01:56) in response to an injection of energy assumed to be in the form of nonthermal electrons. Hard X-ray observations from RHESSI were used to track the evolution of the parameters of the nonthermal electron distribution to reveal the total power contained in flare accelerated electrons. By integrating over the duration of the impulsive phase, the total energy contained in the nonthermal electrons was found to be $>2\times10^{31}$~erg. The response of the lower solar atmosphere was measured in the free-bound EUV continua of \ion{H}{1} (Lyman), \ion{He}{1}, and \ion{He}{2}, plus the emission lines of \ion{He}{2} at 304\AA\ and \ion{H}{1} (Ly$\alpha$) at 1216\AA\ by SDO/EVE, the UV continua at 1600\AA\ and 1700\AA\ by SDO/AIA, and the WL continuum at 4504\AA, 5550\AA, and 6684\AA, along with the \ion{Ca}{2} H line at 3968\AA\ using Hinode/SOT. The summed energy detected by these instruments amounted to $\sim3\times10^{30}$~erg; about 15\% of the total nonthermal energy. The \lya line was found to dominate the measured radiative losses. Parameters of both the driving electron distribution and the resulting chromospheric response are presented in detail to encourage the numerical modelling of flare heating for this event, to determine the depth of the solar atmosphere at which these line and continuum processes originate, and the mechanism(s) responsible for their generation.
\end{abstract}


\section{Introduction}
\label{intro}

First proposed over 40 years ago, the collisional thick-target model (CTTM; \citealt{brow71}) has come to underpin much of our current knowledge of solar flare physics, from the transport of energy from the corona to the chromosphere by means of a beam of electrons, to impulsive hard X-ray (HXR) emission, and the heating of the lower atmosphere. However, it has recently been called into question, and alternative theories are currently being proposed. For example, \cite{flet08} propose that energy is transferred from the corona via Alfv\'{e}n waves, while \cite{brow09} suggest that the electrons are accelerated at the location of the HXR footpoints themselves. \cite{mart12} also recently presented evidence of HXR emission emanating from the same `height' as the corresponding white light (WL) emission during a limb flare, in apparent violation of the CTTM prediction.

It is believed that the bulk of a flare's energy (that deposited by the nonthermal electrons) is radiated via emission emanating from the lower atmosphere, largely in the form of optical continuum radiation \citep{neid89,huds92,huds06}. In addition, the extreme ultra-violet (EUV) component of this radiated emission is believed to be a major energy input into the Earth's upper atmosphere and the geospace environment, heating the thermosphere and generating the ionosphere \citep{lean09,qian10}. It is, therefore, not possible to truly understand the flaring process (including electron acceleration and energy transport) and its consequences without understanding the behavior of the chromosphere during these explosive events.

Simulations have been made by \cite{allr05} using the RADYN code of \cite{carl95,carl97}, which models the chromospheric response to both electron beam heating and backwarming from X-ray and EUV photons. They suggest that chromospheric flare emission is energetically dominated by various recombination (free-bound) continua, in particular the Lyman, Balmer, and Paschen continua of hydrogen, plus the He I and He II continua, as opposed to line (bound-bound) emission. However, definitive observations of free-bound emission during solar flares have been scarce in recent years as most modern space-based instruments have not had the sensitivity, wavelength coverage, or duty cycle required to capture unambiguous continuum enhancements during flares. Understanding how different continua contribute to the overall energy of flares, including the depth of the atmosphere at which they are emitted and the mechanisms by which they are generated, are crucial for testing solar flare models.

Recent observations of solar flares in the Total Solar Irradiance (TSI; \citealt{wood04,wood06}) have placed constraints on the total energy emitted during the largest events (at a few times 10$^{32}$~erg for $>$X10 flares). The authors suggested that about 70\% of this energy comes from wavelengths longer than 270\AA. This is in agreement with \cite{kret10} and \cite{kret11}, who used a superposed epoch analysis of flares over a range of magnitudes. These and other studies conclude that the bulk of a flare's energy is radiated by the WL (``blue'') continuum during the impulsive phase, although they lacked complementary HXR observations to confirm that the energy radiated matched the energy of the nonthermal electrons deemed to be responsible. However, \cite{neid93} concluded that there was sufficient energy in nonthermal electrons above 48~keV to drive the associated optical emission for a WL flare that occurred on 1989 March 7. Similarly, \cite{zhar07} used time-dependent beam parameters from HXR observations in an attempt to recreate \ha observations using hydrodynamic modelling; reasonable agreement was found. More recently, \cite{flet13} found that the electron beam power derived from HXR observations (assuming thick-target collisions) could account for the UV, EUV, and soft X-ray (SXR) components of flare ribbons during an M-class flare, as long as the low energy cutoff was set at 4--5~keV.

\begin{figure}[!t]
\begin{center}
\includegraphics[width=0.5\textwidth]{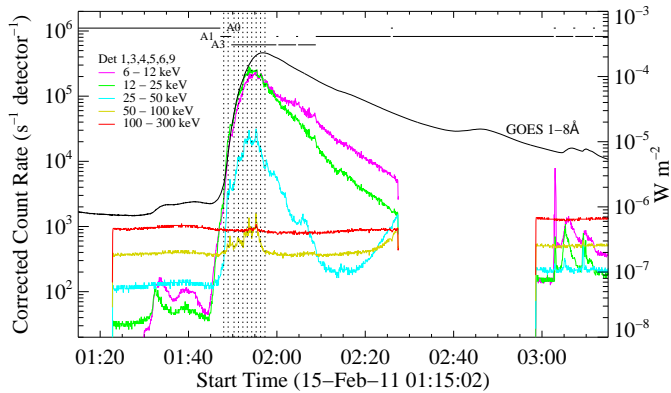}
\caption{RHESSI count rate time profiles (corrected for changes in attenuator states) in 5 energy bands (6--12, 12--25, 25--50, 50--100, and 100--300~keV) during the 2011 February 15 X2.2 flare. Also plotted is the GOES 1--8\AA\ lightcurve. The vertical dotted lines denote the ten 60~s time intervals over which spectra were obtained.}
\label{goes_hsi}
\end{center}
\end{figure}

This paper examines how the energy deposited in the lower solar atmosphere by energetic electrons during the impulsive phase of an X-class solar flare, as determined from HXR observations, gets redistributed across the optical, UV, and EUV portions of the spectrum throughout the duration of the event, with particular emphasis on continuum processes. The details of the parent electron distribution are provided as input parameters for numerical simulations that can model the response of the lower solar atmosphere to an injection of energy. The output generated by such models can then be directly compared with the observed chromospheric response. Such comparisons provide a better understanding of the dominant radiation processes during flares \citep{huds10} and information on the depth at which various line and continuum emission processes take place, and the mechanism(s) responsible for generating them. Section~\ref{data_anal} describes how the data from each instrument were analyzed. The findings from this analysis are presented in Section~\ref{results}, while their implications are presented and discussed in Section~\ref{conc}.

\section{Observations and Data Analysis}
\label{data_anal}

On 2011 February 15 the first X-class flare of Solar Cycle 24, an X2.2 flare that began at 01:44~UT (SOL2011-02-15T01:56), occurred in NOAA Active Region 11158 (Solar X = 417\arcsec, Solar Y = -433\arcsec). The X-ray lightcurves of the event are shown in Figure~\ref{goes_hsi}. This event and its parent active region were extensively observed by a broad range of space-based solar observatories, resulting in over 30 published papers to date. It is this extensive coverage that makes it an ideal candidate for studying the atmospheric response to an injection of energy, assumed to be in the form of a beam of nonthermal electrons. Data from the Ramaty High-Energy Solar Spectroscopic Imager (RHESSI; \citealt{lin02}) were used to measure the energy flux of the electrons assumed to be responsible for driving the enhanced chromospheric emission (Section~\ref{rhessi}). The EUV Variability Experiment (EVE; \citealt{wood12}) instrument onboard the Solar Dynamics Observatory (SDO; \citealt{pesn12}) provided full-disk observations of the free-bound EUV continua (Lyman, \ion{He}{1}, and \ion{He}{2}), as well as the chromospheric \ion{He}{2} 304\AA\ and \lya 1216\AA\ lines at 10~s cadence (Section~\ref{eve}), while the Atmospheric Imaging Assembly (AIA; \citealt{leme12}), also on SDO, provided measurements of the UV continuum in its 1600\AA\ and 1700\AA\ passbands at 24~s cadence (Section~\ref{aia}). Finally, the optical continuum at 4504\AA, 5550\AA, and 6684\AA, as well as the \ion{Ca}{2} H line at 3968\AA, were recorded by the Solar Optical Telescope (SOT; \citealt{tsun08}) onboard Hinode at 20~s cadence. In this section the data analysis techniques for each instrument are described. 

\begin{table*}[!t]
\begin{center}
\small
\caption{\textsc{\small{Mean$\pm$1$\sigma$ Values of the Nonthermal Electron Distribution Parameters}}}
\label{hsi_beam_params}
\begin{threeparttable}
\begin{tabular}{ccccccc} 
\tableline
\tableline
Time UT		&Total Electron 				&Spectral  	&Low Energy		&Nonthermal Power 			&Footpoint				&Electron Energy Flux		\\ 
($\pm30s$)	&Rate (electrons s$^{-1}$)	&Index		&Cutoff (keV)		&(\ergs)					&Area (cm$^{2}$) 			&(\ergcms)				\\ 
\tableline
01:47:58		&$5.7\pm2.6\times10^{34}$	&$7.2\pm0.5$	&$<25.3\pm3.4$	&$>2.7\pm1.1\times10^{27}$	&$0.8\times10^{17}$\tnote{a}	&$>3.3\pm1.4\times10^{10}$	\\
01:48:58		&$3.0\pm0.7\times10^{35}$	&$6.1\pm0.1$	&$<25.9\pm1.9$	&$>1.5\pm0.4\times10^{28}$	&$1.2\times10^{17}$\tnote{a}	&$>1.3\pm0.3\times10^{11}$	\\
01:50:18		&$2.4\pm1.0\times10^{36}$	&$7.2\pm0.1$	&$<21.9\pm1.6$	&$>9.9\pm3.7\times10^{28}$	&$2.1\times10^{17}$		&$>4.7\pm1.7\times10^{11}$	\\
01:51:18		&$2.1\pm1.2\times10^{36}$	&$6.9\pm0.2$	&$<23.2\pm1.1$	&$>9.5\pm4.7\times10^{28}$	&$3.2\times10^{17}$		&$>3.0\pm1.5\times10^{11}$	\\
01:52:18		&$1.5\pm1.0\times10^{36}$	&$6.3\pm0.3$	&$<22.3\pm1.6$	&$>6.6\pm3.9\times10^{28}$	&$3.4\times10^{17}$		&$>2.0\pm1.2\times10^{11}$	\\
01:53:18		&$7.2\pm3.2\times10^{35}$	&$5.4\pm0.2$	&$<22.9\pm1.0$	&$>3.4\pm1.5\times10^{28}$	&$4.1\times10^{17}$		&$>8.3\pm3.6\times10^{10}$	\\
01:54:18		&$2.7\pm1.2\times10^{35}$	&$4.8\pm0.2$	&$<22.4\pm1.3$	&$>1.3\pm0.6\times10^{28}$	&$4.2\times10^{17}$		&$>3.1\pm1.4\times10^{10}$	\\
01:55:18		&$3.0\pm1.6\times10^{35}$	&$4.6\pm0.2$	&$<21.8\pm0.9$	&$>1.5\pm0.7\times10^{28}$	&$4.2\times10^{17}$		&$>3.5\pm1.7\times10^{10}$	\\
01:56:18		&$1.5\pm1.0\times10^{35}$	&$5.0\pm0.4$	&$<22.8\pm2.0$	&$>7.5\pm4.5\times10^{27}$	&$2.8\times10^{17}$		&$>2.7\pm1.6\times10^{10}$	\\
\tableline
\normalsize
\end{tabular}
\begin{tablenotes}
\item[a]{Estimated}
\end{tablenotes}
\end{threeparttable}
\end{center}
\end{table*}

\subsection{RHESSI}
\label{rhessi}
High-energy emissions are the most direct signature of particle acceleration during the impulsive phase of a solar flare. Accelerated electrons colliding with the ambient solar atmosphere produce HXR and $\gamma$-ray free-free (bremsstrahlung) continuum emission. RHESSI was designed to provide high-resolution imaging and spectroscopy of these emissions. If the HXR emission is chromospheric, it is reasonable to assume thick-target interactions, and under this assumption the parameters of the electron spectrum can be deduced from the measured photon spectrum. The total power contained in the electron distribution ($P_{nth}$) can be calculated using:
\begin{equation}
P_{nth}(E \geq E_C) = \int_{E_C}^{\infty} E F(E) dE~~{\mbox{\ergs}},
\label{p_nth}
\end{equation}
where $F(E)$ is the electron distribution in the form of a power-law given by $AE^{-\delta}$~(electrons~s$^{-1}$~keV$^{-1}$). The normalization factor, $A$, is proportional to the total injected electron rate, $E$ is the electron energy, $E_C$ is the low-energy cutoff, and $\delta$ the spectral index of the electron distribution. The expression for $P_{nth}$ therefore becomes:

\begin{equation}
\label{e_dist_func}
P_{nth}(E \geq E_C) = \frac{\kappa{_E}A}{(\delta-2)}E_{C}^{(2-\delta)}~~{\mbox{\ergs}},
\end{equation}
where $\kappa_E$ is the conversion factor from keV to erg ($1.6\times10^{-9}$). 

While RHESSI can be used to deduce the electron spectral index very accurately, there are large uncertainties associated with the values determined for the low-energy cutoff because the dominance of thermal emission at low energies makes it harder to identify \citep{holm03,irel13}. This leads to a lower limit on the total energy contained in the electron distribution, due to the value of $E_C$ being taken as an upper limit.

\begin{figure}[!t]
\begin{center}
\includegraphics[width=0.5\textwidth]{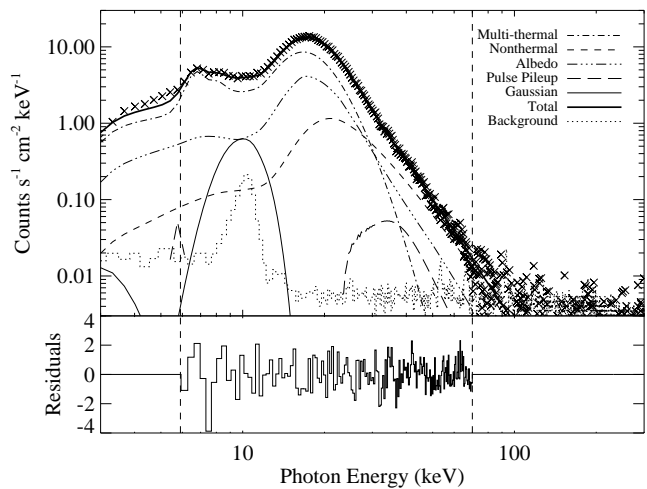}
\caption{RHESSI count spectrum from detector 4 integrated from 01:51:48--01:52:48~UT on 2011 February 15 while the thick attenuators were in place (A3 state). Overlaid are the fits to the various components that comprise the total fit (see legend) between 6 and 70~keV. The bottom panel shows the normalized residuals to the total fit.}
\label{hsi_spec}
\end{center}
\end{figure}

The impulsive phase of the 2011 February 15 event lasted for $\sim$10 minutes. RHESSI spectra were compiled for ten 60~s integrations (denoted by the vertical dashed lines in Figure~\ref{goes_hsi}) for each of the nine germanium detectors individually, and were fitted with the sum of a thermal component that dominated at low energies and a collisional thick-target model with a power-law electron spectrum at higher energies ({\sc thick2\_vnorm}). The thermal component was assumed to be isothermal for the rise phase and decay phase of the event. However, for the 5 time bins around the peak of the flare (01:51:48--01:56:48 UT) the count spectra were better fit with a multi-thermal component with a DEM of power law in log T (see sample RHESSI spectrum in Figure~\ref{hsi_spec}). A similar technique was employed by \cite{casp10}. Taking the mean and standard deviation of each fit parameter across individual detectors currently provides the best estimates of the spectral parameters and their uncertainties ({\it c.f.} \citealt{mill09}). Compiling spectra for each of RHESSI's  detectors individually also allows the most up-to-date albedo \citep{kont06}, pulse pileup, and gain offset estimates currently available in the OSPEX software package to be used.

\begin{figure}[!t]
\begin{center}
\includegraphics[width=0.5\textwidth]{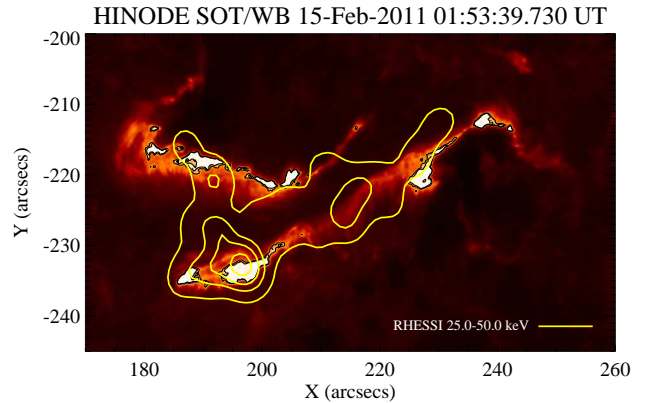}
\caption{Image of the flare ribbons in the \ion{Ca}{2} H line taken at 01:53~UT by Hinode/SOT. Black contours mark the 80\% level of peak intensity. The yellow contours denote the 30\%, 50\%, and 80\% levels of the associated RHESSI 25--50~keV image taken around the same time.}
\label{hsi_sot}
\end{center}
\end{figure}

\begin{figure*}[!t]
\begin{center}
\includegraphics[width=0.7\textwidth]{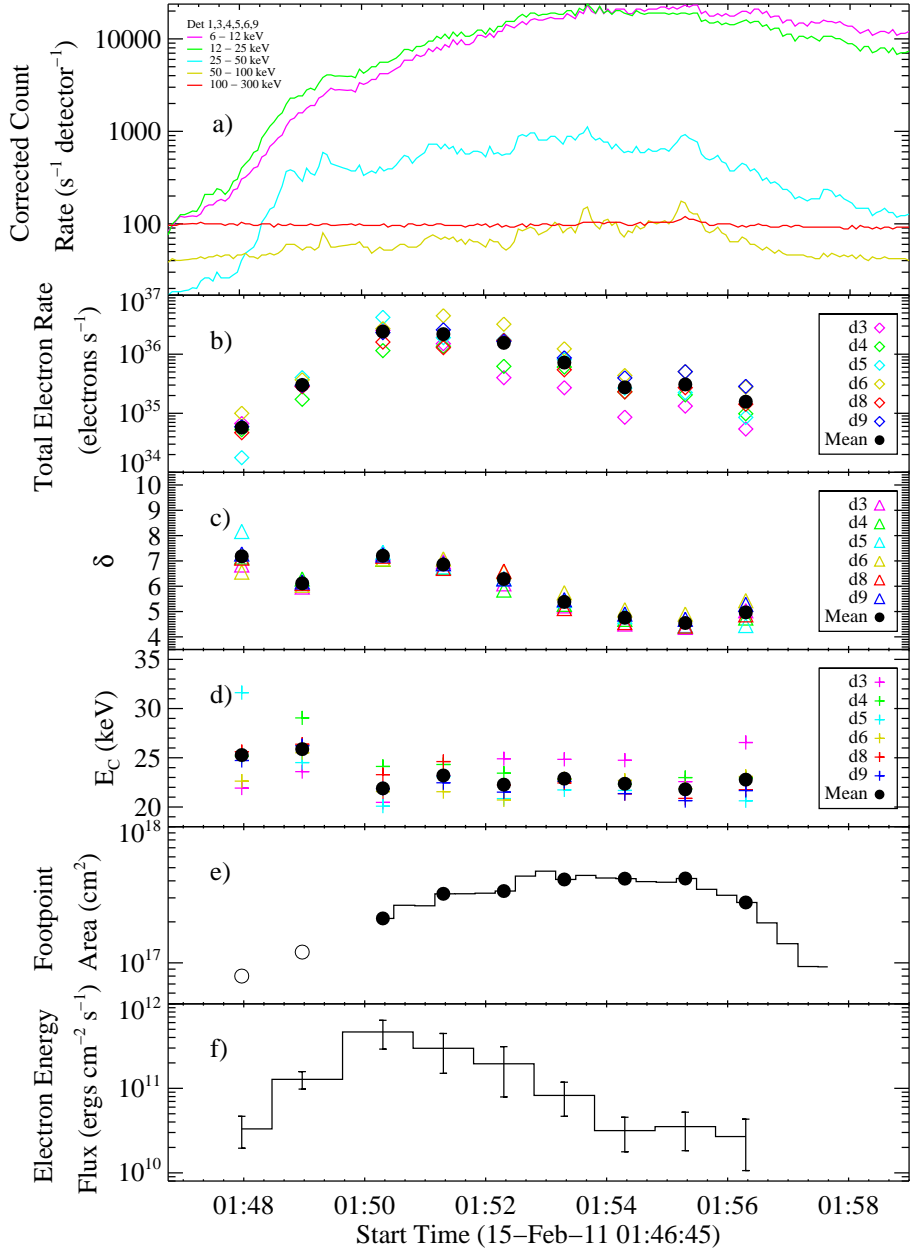}
\caption{Electron distribution parameters derived from RHESSI HXR spectra. $a)$ Corrected count rate lightcurves in five energy bands; $b)$ total electron rate derived from six individual detectors (colored diamonds); $c)$ electron spectral index, $\delta$, for six individual detectors (colored triangles); $d)$ low-energy cutoff, $E_C$, for six individual detectors (colored plus signs). The mean value of the electron rate, $\delta$, and $E_C$ across all detectors at each time is plotted as a filled black circle; $e)$ footpoint areas derived from SOT \ion{Ca}{2} H images. The filled black dots denote the SOT areas measured closest to the center of each 60~s RHESSI time interval. SOT was not observing before $\sim$01:50~UT and so the first two data points (empty circles) are estimates; $f)$ electron energy flux calculated from the mean power at each time interval divided by the footpoint areas from SOT.} 
\label{ebeam_params}
\end{center}
\end{figure*}

Dividing the power contained in the electrons by the footpoint area over which their energy is deposited allows the electron energy flux (in \ergcms) to be determined. It is assumed that RHESSI HXR images of flare footpoints are largely unresolved, similar to flare ribbons \citep{denn09}, in contrast to WL images of flare ribbons which are resolved \citep{kruc11}. Figure~\ref{hsi_sot} shows an image of the flare ribbons for this event as observed by SOT in the \ion{Ca}{2} H line at 01:53:39~UT. Overlaid are the contours of the concurrent 25--50~keV emission observed by RHESSI, reconstructed using the CLEAN algorithm (detectors 2--8, with a CLEAN beam width factor of 2). SOT images were manually shifted by Solar X = +28\arcsec\ and Solar Y = +22\arcsec\ to align with those from RHESSI. It is immediately apparent that the area obtained from RHESSI images would be an overestimate of the true region over which the beam energy was deposited. Therefore, the pixels with intensities $>$80\% of the peak emission in each \ion{Ca}{2} H image (black contours around the brightest emission in Figure~\ref{hsi_sot}) were summed to give a more realistic estimate of the footpoint area for use in determining the electron beam flux density as a function of time (see below). 

\begin{figure*}[!t]
\begin{center}
\includegraphics[width=\textwidth]{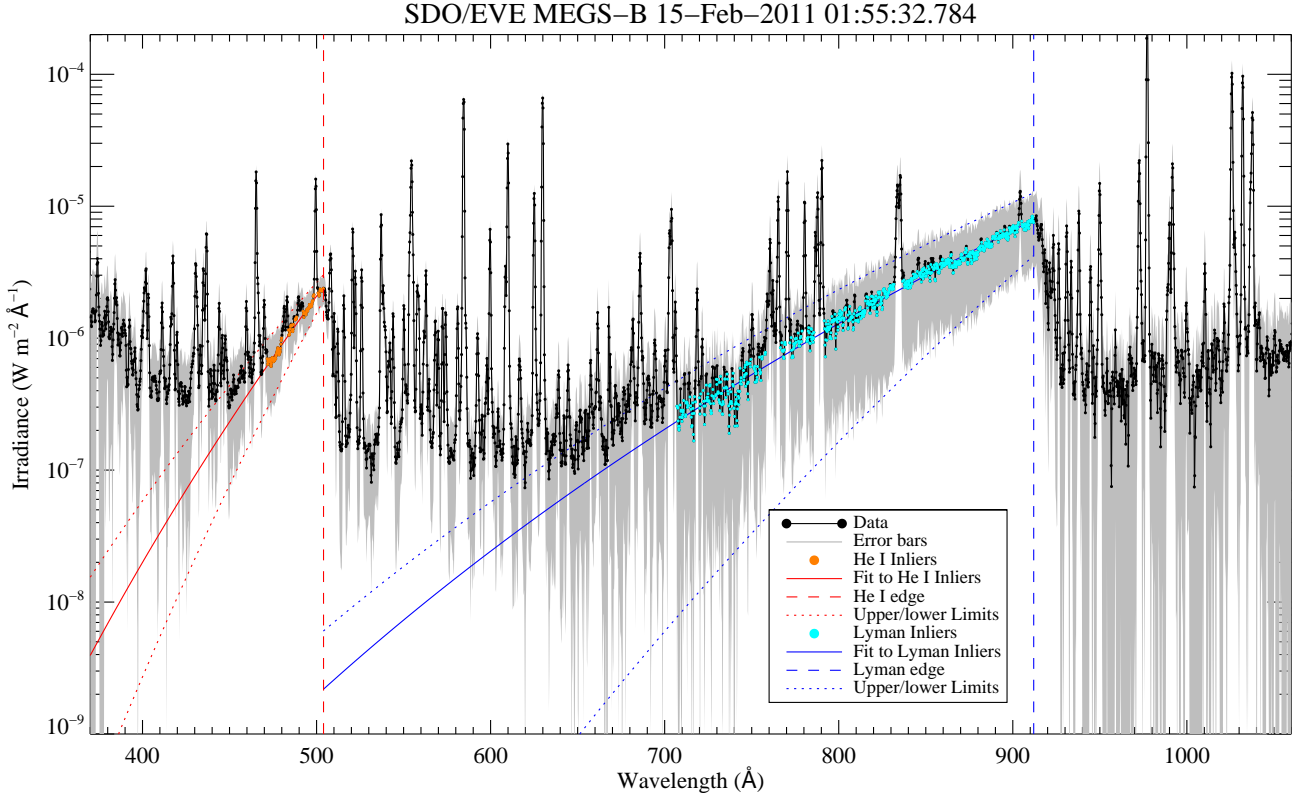}
\caption{Full-Sun SDO/EVE MEGS-B spectrum from 370--1060\AA\ taken at the SXR peak of the flare. The grey shaded area denotes the range of uncertainties in the irradiance, as obtained from the EVE database. We believe these to be generous upper limits on the relative systematic errors of these measurements. Overlaid are the fits to the \ion{He}{1} (red) and Lyman (blue) continua blueward of the recombination edges at 504\AA\ and 912\AA\ (vertical dashed lines), respectively, using the RANSAC method (see Appendix). The orange and cyan data points denote the inliers attributed to the \ion{He}{1} and Lyman continua, respectively. The dotted lines are fits to the upper and lower limits over the same wavelength ranges for each of the two continua.}
\label{eve_megsb_cont_fits}
\end{center}
\end{figure*}

Figure~\ref{ebeam_params}a shows the count rate of HXR emission (corrected for changes in attenuator states) in five energy bands as observed by RHESSI. Panels b--d show the evolution of the total electron rate, spectral index, and low-energy cutoff, respectively, as determined from the thick-target model fitted to the RHESSI spectra over the course of the flare for detectors 3, 4, 5, 6, 8, and 9 individually (colored symbols). The solid black dots denote the mean value of each parameter by averaging over these six detectors at each time (these values are also listed in Table~\ref{hsi_beam_params} along with their 1$\sigma$ uncertainties). Figure~\ref{ebeam_params}b reveals that the mean total electron number rate increased by almost 2 orders of magnitude during the first 3 minutes of the event, after which it began to decrease. The low-energy cutoff (upper limit; Figure~\ref{ebeam_params}d) stayed between 21~keV and 26~keV for the duration of the event. Also plotted are the footpoint areas from SOT (Figure~\ref{ebeam_params}e). SOT was not observing at the beginning of the impulsive phase, so the first two data points were estimated by extrapolating back in time linearly from the first two measured data points. The total area derived from SOT images was found to increase until the end of the impulsive phase, while the corresponding  area returned from RHESSI images remained fairly constant over the course of the flare ($\sim10^{18}$~cm$^{2}$). Figure~\ref{ebeam_params} (bottom panel) shows the evolution of the electron energy flux (in \ergcms) by taking the mean power derived at each time interval (from Equation~\ref{p_nth}) and dividing by the associated footpoint area from SOT. These values are also listed in Table~\ref{hsi_beam_params}. The electron energy flux appeared to mimic the total electron rate, peaking 3 minutes into the event, after which it decreased back to its initial value. Error bars denote the 1$\sigma$ standard deviations of the total power across all six detectors, although in actuality, the values quoted ought to be considered as lower limits given that the low-energy cutoff values are upper limits, as are the footpoint areas, assuming a filling factor of unity.

\subsection{SDO/EVE}
\label{eve}

SDO/EVE acquires full-disk EUV spectra every 10~s over the 65--370\AA\ wavelength range using the MEGS-A (Multiple EUV Grating Spectrographs) component with a near 100\% duty cycle. The MEGS-B component (370--1060\AA), which covers the He I and Lyman continua, and the MEGS-P broadband diode centred on the \lya line at 1216\AA, have a reduced duty cycle due to unforeseen instrument degradation. Although EVE  was primarily designed to monitor changes in the Sun's EUV irradiance over multiple timescales, several studies have shown that its data can be utilized to probe the physical parameters of solar flare plasmas at high cadence (e.g. \citealt{huds11}). \cite{mill12b} were able to track the evolution of flare densities at high ($>$10~MK) temperatures from pairs of density-sensitive \ion{Fe}{21} lines. \cite{kenn13} used a Markov-Chain Monte Carlo method to reconstruct flare Differential Emission Measures (DEM) from EVE data. Concurrent observations from AIA revealed that these DEMs were representative of the flaring chromosphere.

On 2013 December 20, Version 4 of the EVE data were released, marking a significant advancement over previous versions, with particular emphasis on correcting the long-term degradation of the MEGS-B instrument. This section describes how EVE data were used to quantify the energy emitted in EUV lines and continua formed in the chromosphere over the course of the flare.

\begin{figure}[!t]
\begin{center}
\includegraphics[width=0.5\textwidth]{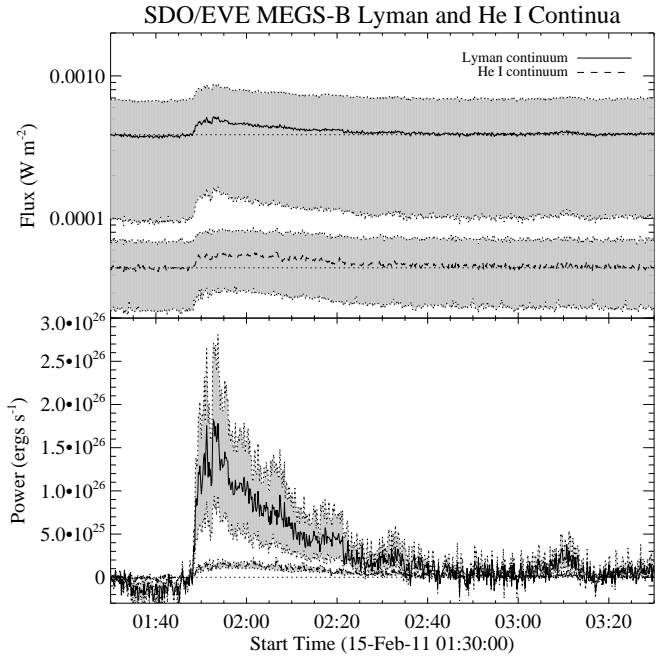}
\caption{Time profiles of the \ion{He}{1} (dashed curve) and Lyman (solid curve) continua compiled by integrating under the fits to the spectra shown in Figure~\ref{eve_megsb_cont_fits}, before (top panel) and after (bottom panel) background subtraction. Grey shaded areas are bounded by fits to the upper and lower limits of each continuum, using the uncertainty estimates from the EVE database.}
\label{eve_megsb_cont_ltc}
\end{center}
\end{figure}

\begin{figure}[!t]
\begin{center}
\includegraphics[width=0.5\textwidth]{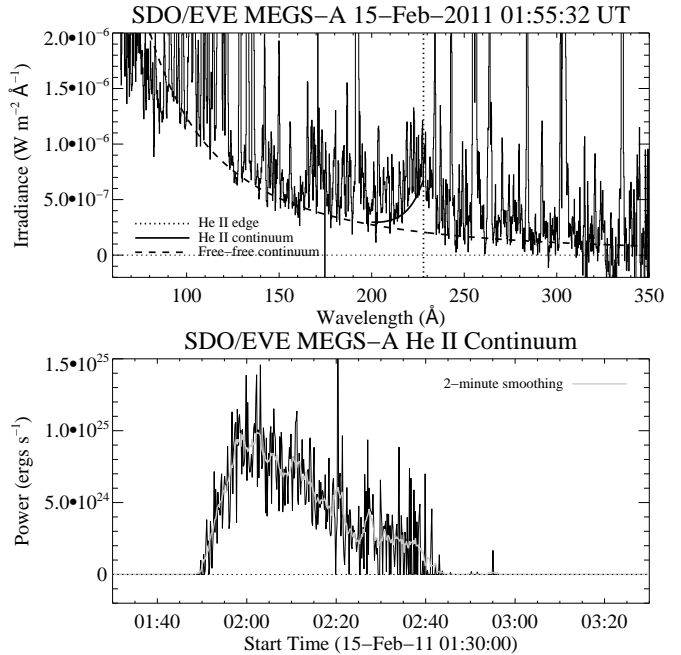}
\caption{Top panel: A complete EVE MEGS-A spectrum taken near the peak of the 2011 February 15 flare having subtracted out a pre-flare profile. The \ion{He}{2} recombination edge at 228\AA\ is shown by the vertical dashed line. Fits to the free-free and free-bound \ion{He}{2} continua are shown as dotted and solid curves, respectively. Bottom panel: Time profile of the \ion{He}{2} continuum taken from the integral of the fit in the top panel over the course of the flare, after subtracting out a pre-flare spectrum and the fit to the underlying free-free continuum. Overlaid in grey is a 2-minute smoothed profile for clarity.}
\label{eve_heii_cont_ltc}
\end{center}
\end{figure}

\subsubsection{EUV Continua}
\label{euv_continua}

\cite{mill12a} initially presented observations from EVE that showed unambiguous, spectrally and temporally-resolved detections of enhanced free-bound (and free-free) continua during the flare presented here. That study was conducted using Version 2 of EVE data, and an {\it ad hoc} method was employed to obtain time profiles of the free-bound emission. Line-free portions of the EVE spectra based on a synthetic flare spectrum from CHIANTI were averaged and fit with an exponential function. In this revised analysis, a more robust technique was developed and applied to Version 4 data along with the associated uncertainties as provided by the EVE database.

As the \ion{He}{1} and Lyman continua (MEGS-B) are both evident in quiet-Sun spectra, they were both fit with a power-law function in wavelength using the RANdom Sample Consensus (RANSAC) method (\citealt{fisc81}; see Appendix for details). RANSAC is a method to determine the parameters of a function used to represent an observational dataset that includes outliers. This approach is best suited for fitting the continua in the MEGS-B spectral range as it treats the emission lines superimposed on the continua, which vary in intensity throughout a flare, as outliers. A straightforward, least-squares fit to all the data points in this part of the spectrum would be biased by these lines and would not give a true representation of the continuum alone. The inliers for the \ion{He}{1} continuum were fitted between 470\AA\ and the recombination edge at 504\AA, and then extrapolated down to 370\AA, while the Lyman continuum inliers were fitted from 700\AA\ to the Lyman edge at 912\AA, and then extrapolated down to the \ion{He}{1} edge at 504\AA\ (see Figure~\ref{eve_megsb_cont_fits}). The uncertainties on the EVE MEGS-B data are shown in Figure~\ref{eve_megsb_cont_fits} in grey. The upper and lower boundaries to each continua were also fit using the RANSAC method. By integrating under these fits, lightcurves of the (full-Sun) free-bound continua and their uncertainties were derived (top panel of Figure~\ref{eve_megsb_cont_ltc}). Subtracting out a pre-flare background and converting from flux values (W~m$^{-2}$) to power (\ergs) revealed how the energy radiated by these continua changed as a function of time during the flare (bottom panel of Figure~\ref{eve_megsb_cont_ltc}). The Lyman continuum had a peak radiative loss rate of $\sim2\times10^{26}$~\ergs, while the \ion{He}{1} continuum peaked around $2\times10^{25}$~\ergs.

The \ion{He}{2} continuum (MEGS-A), on the other hand, is not immediately evident outside of flaring conditions as it is inherently weak. It also competes with the free-free (thermal bremsstrahlung) continuum, which is also brighter during flares, and therefore required a different fitting technique. \cite{mill13} used EVE MEGS-A data (Version 3) to quantify the amount of free-free emission (and free-bound in the case of the 211\AA\ channel) that contributed to each of the EUV passbands on AIA. They were able to fit an exponential function to the lower envelope of the EVE spectra by differentiating twice with respect to wavelength to identify the local minima (essentially, the `turning points' between emission lines). The same technique was repeated here, but to Version 4 of the EVE data. In the top panel of Figure~\ref{eve_heii_cont_ltc} an EVE MEGS-A spectrum taken at the SXR peak of the flare after subtracting out a pre-flare profile is shown. The dotted line denotes the fit to the free-free continuum across the entire MEGS-A wavelength range, while the solid line shows the corresponding fit to the \ion{He}{2} free-bound continuum from 200\AA\ up to the recombination edge at 228\AA\ (vertical dashed line). In the bottom panel the time profile of the integral under this fit is shown. There is a large amount of scatter in the lightcurve due to noise of the background-subtracted data above the free-free continuum. A 12-point (2 minute) smoothing has been applied to the lightcurve for clarity. The \ion{He}{2} continuum peaked at $\sim1\times10^{25}$~\ergs.

\begin{figure}[!t]
\begin{center}
\includegraphics[width=0.5\textwidth]{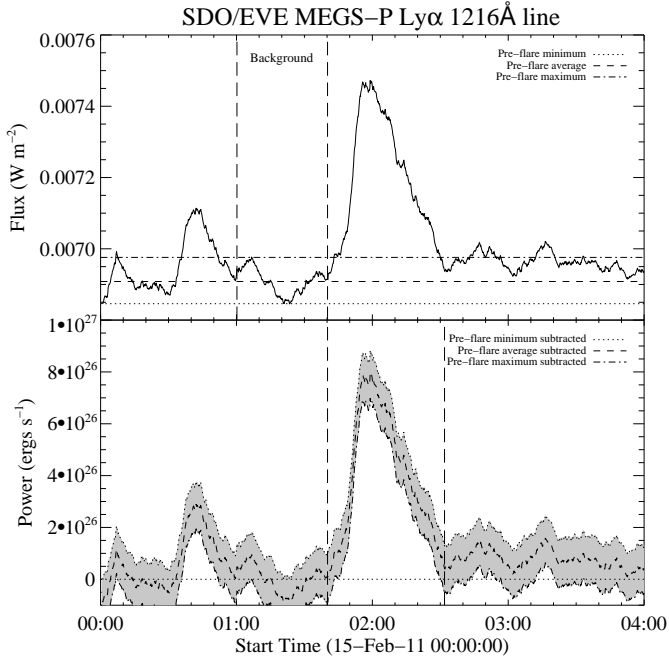}
\caption{Top panel: Full-disk lightcurves in the Ly$\alpha$ line from EVE MEGS-P. Pre-flare irradiance levels were taken from the minimum, maximum, and average between 01:00--01:40~UT (vertical long-dashed lines). Bottom panel: Flare lightcurves in terms of energy having subtracted both a pre-flare minimum (dotted line), averaged (dashed line), and maximum (dot-dashed line) background, excluding the separate C-class flare event at about 00:40 UT. The total energy emitted by the Ly$\alpha$ was found by integrating between the two vertical long-dashed lines.}
\label{lya_ltc}
\end{center}
\end{figure}

\subsubsection{EUV Lines}
\label{euv_lines}

The two strongest lines in the EVE spectral range are the chromospheric \ion{He}{2} 304\AA\ line (in MEGS-A) and the \ion{H}{1} \lya line at 1216\AA\ (covered by the MEGS-P diode). Both are believed to be significant radiators of energy deposited in the chromosphere during a solar flare, while \lya emission is also a prominent driver of terrestrial atmospheric variations \citep{tobi00}. Despite its importance, there have been relatively few \lya flares reported in the literature (see references in \citealt{kret13}). Although the MEGS-P response function is 100\AA\ wide, more than 99\% of the detected emission is solar \lya after accounting for non-solar sources (e.g., geocoronal emission; \citealt{cros04,wood10}). In addition, the \lya channel is a photodiode versus a CCD for the other MEGS channels, so the high-energy particles in the SDO orbit cause spikes in the \lya data time series as opposed to just affecting a pixel or two on the CCD sensors. The data processing algorithms attempt to make a correction for these spikes in the data time series, but not all of the spikes are successfully removed. Consequently, the EVE \lya measurements have more noise in their time series. As such, the largest uncertainty in determining the flare-related energy emitted by \lya is due to large fluctuations in the background, as can be seen in the full-disk lightcurves in the top panel of Figure~\ref{lya_ltc}. Therefore, the pre-flare minimum (dotted line), average (dashed line), and maximum (dot-dashed line) levels between 01:00--01:40~UT were taken as background levels with which to compute a likely range of \lya energies (bottom panel of Figure~\ref{lya_ltc}). Around the peak of the event the \lya line was emitting $8\times10^{26}$~\ergs, while the \lya flux returned to its pre-flare level around 02:30~UT.

\begin{figure}[!t]
\begin{center}
\includegraphics[width=0.5\textwidth]{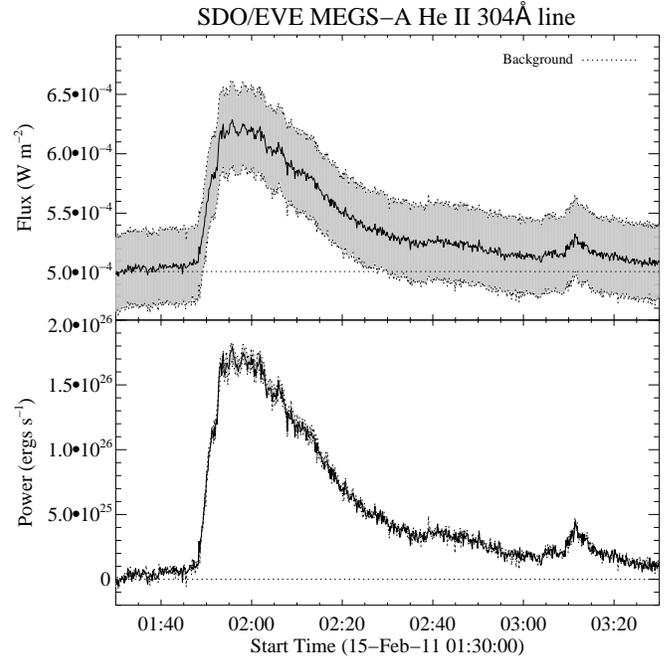}
\caption{Top panel: Full-disk lightcurves in the \ion{He}{2} 304\AA\ line from EVE MEGS-A. Pre-flare irradiance levels were taken from a 5 minute average between 01:00--01:05~UT. Bottom panel: Flare lightcurves in terms of energy having subtracted a pre-flare background.}
\label{heii_ltc}
\end{center}
\end{figure}

The \ion{He}{2} 304\AA\ line is observed in the MEGS-A channel. At each time step, the line was fitted with a Gaussian profile and integrated over 10 wavelength bins (302.9\AA--304.9\AA) to obtain the line flux. The time profile exhibited a relatively stable pre-flare background (top panel of Figure~\ref{heii_ltc}). After subtracting out this pre-flare value and converting to energy units, it was found that the \ion{He}{2} line emission remained elevated for over 90 minutes after the main flaring event, twice as long as the \lya line. Its peak radiative loss rate was $\sim1.7\times10^{26}$~\ergs. Note that during an M2 flare \cite{kret13} found the \lya and \ion{He}{2} 304\AA\ time profiles, using data from PROBA2/LYRA and SOHO/SEM, respectively, to be almost identical.

\subsection{SDO/AIA}
\label{aia}

SDO/AIA produces high-resolution, full-disk images of the Sun in ten wavelength bands. Seven of these passbands are tuned to observe predominantly coronal emission in the EUV at 12~s cadence, while two were designed to detect chromospheric emission in the ultraviolet (UV; 1600\AA\ and 1700\AA) at 24~s cadence. In quiet-Sun conditions, the 1700\AA\ channel samples the UV continuum near the temperature minimum in the photosphere, while the 1600\AA\ channel covers the upper photosphere and transition region, and also includes emission from a \ion{C}{4} line.

\begin{figure}[!t]
\begin{center}
\includegraphics[width=0.5\textwidth]{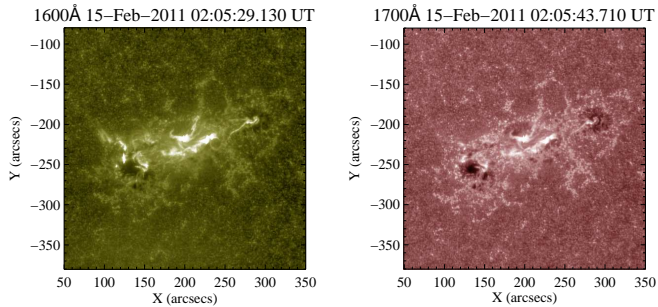}
\caption{Images of the 2011 February 15 flare ribbons in the two UV bands on AIA. Left: 1600\AA. Right 1700\AA.}
\label{aia_maps}
\end{center}
\end{figure}

\begin{figure}[!t]
\begin{center}
\includegraphics[width=0.5\textwidth]{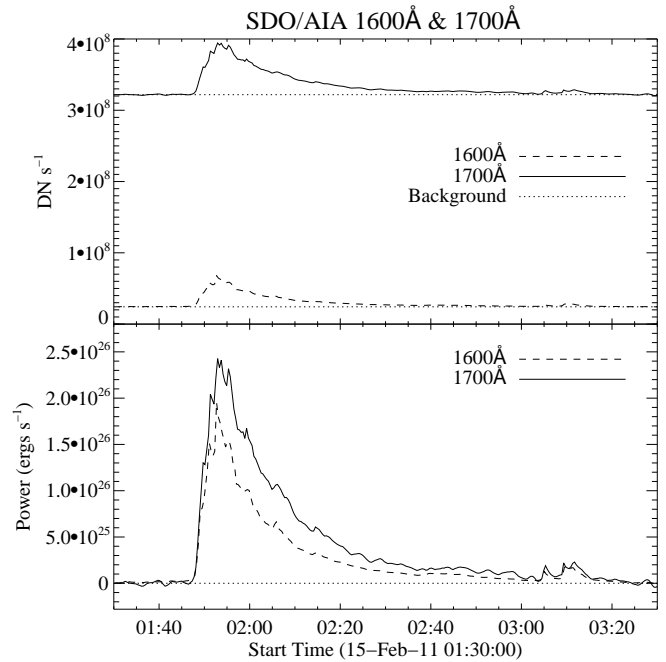}
\caption{Lightcurves of the total UV emission observed in each of the two AIA channels. Top panel: Raw counts in DN s$^{-1}$. Bottom panel: pre-flare subtracted time profiles in units of energy.}
\label{aia_ltc}
\end{center}
\end{figure}

Figure~\ref{aia_maps} shows the flare ribbons for the 2011 February 15 event in both the 1600\AA\ and 1700\AA\ channels. These channels both suffered from saturation and bleeding around the peak of this event. However, no loss of counts occurs if the total rate is determined over an area large enough to contain all of the saturation and bleeding. Thus, by integrating over the areas shown the total number of photons incident on the detectors could be reliably determined (Dr. Paul Boerner, 2014; private communication). Lightcurves of this integrated emission are shown in the top panel of Figure~\ref{aia_ltc} (in DN s$^{-1}$). Subtracting out a pre-flare background and convolving these profiles with the response functions for each channel (using Version 4 of the response functions in the {\sc aia\_get\_response.pro} routine in SSWIDL), the energy radiated as a function of time could be estimated (bottom panel of Figure~\ref{aia_ltc}). From the resulting lightcurves shown in the top panel of Figure~\ref{aia_ltc}, there are no obvious signs of saturation in the integrated profiles. The time profile of the 1700\AA\ emission peaked at $2.5\times10^{26}$~\ergs, while the 1600\AA\ channel emitted a maximum of $2\times10^{26}$~\ergs.

\subsection{Hinode/SOT}
\label{sot}
 
The \textsl{Hinode}/SOT Broadband Filtergram Imager (BFI) images the Sun in the WL continuum using three narrow channels, red ($6684\pm2$)\AA, green ($5550.5\pm2$)\AA, and blue ($4504.5\pm2$)\AA, collectively termed RGB, as well as taking observations in the \ion{Ca}{2} H line ($3968.5\pm1.5$)\AA, which is measured around the line centre \citep{suem08}. During flare mode, triggered by the X-Ray Telescope (XRT), also on board \textsl{Hinode}, the SOT offers excellent spatial resolution of 0.108\arcsec\ with a typical cadence of $\sim$20~s. While several  studies of white light flares (WLFs) using SOT G-band images have been performed (e.g., \citealt{isob07}), to our knowledge there have only been two investigations of the SOT WL continuum data to date: \citet{wata13} and \citet{kerr14}. Note that whilst \cite{wata13} focused on the temperature of the WL emission in the 2012 January 27 solar flare, rather than the energetics, their results are largely consistent with \cite{kerr14}.

\begin{figure}
\begin{center}
\includegraphics[width = 0.5\textwidth]{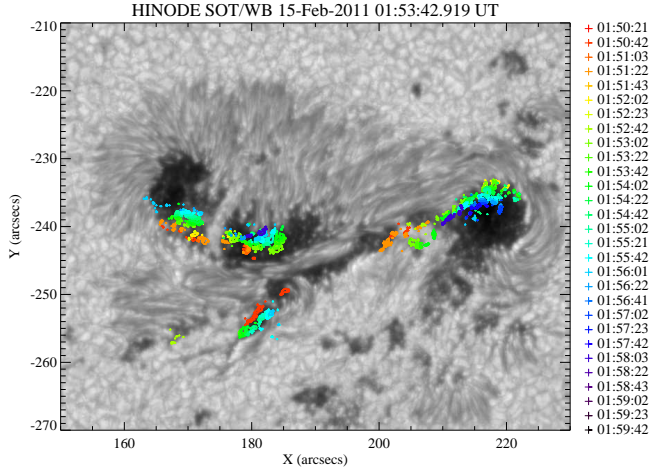}
\caption{Hinode/SOT continuum image for 2011 February 15 flare with the location of the white light footpoints at various times overlaid.}
\label{sot_wl_image}
\end{center}
\end{figure}

SOT observed the 2011 February 15 solar flare from $\sim$01:50UT to $\sim$02:00UT, in the RGB continua and the \ion{Ca}{2} H line. The RGB emission from this flare was investigated by \cite{kerr14}, the results of which are summarised here in the context of the overall energetics of the flare, along with the addition of the \ion{Ca}{2} H line energetics. Data were processed using the \texttt{fg$\_$prep} routine, which corrected for dark currents, flat fields, and removed `hot pixels', and each frame was divided by the exposure time to give intensity measured in DN~s$^{-1}$~px$^{-1}$. Images were co-aligned, with the frame at 01:52:42~UT as a reference. A manual correction for large shifts was performed by overlaying the image contours and shifting the images until they lined up with the pore at Solar X = 225\arcsec, Solar Y = -215\arcsec\ (see Figure~\ref{sot_wl_image}). Any remaining misalignment was removed using the \textsl{Hinode} cross-correlation routine, \texttt{fg$\_$rigid$\_$align}. Conversion from DN~s$^{-1}$~px$^{-1}$ to W~cm$^{-2}$~sr$^{-1}$~\AA$^{-1}$ was done using conversion factors provided by Dr T. Tarbell (2012, private communication). These conversion factors were calculated using the average solar disk spectrum given by the Brault \& Neckel (1987) Spectral Atlas available from the Hamburg Observatory FTP site \citep{neck99}, the filter response of the SOT channels, and the observed quiet Sun intensities measured on 2011 February 14 (c.f. \citealt{kerr14}). These factors are listed in Table~\ref{table:conv_factors}.

\cite{kerr14} made model-dependent estimation of the energetics of the full WL spectrum. Under the assumption of blackbody emission the authors found a temperature increase of $\sim$200~K and an instantaneous power emitted by the newly-brightened sources of $10^{26}$~\ergs\ ($\sim10^{28}$~erg over the duration of the event). Assuming hydrogen recombination emission from an optically thin slab led to a temperature in the range 5,500--25,000~K with an instantaneous excess power of $10^{27}$~\ergs\ ($\sim10^{29}$~erg in total).
 
\begin{table*}
\begin{center}
\caption{\textsc{\small{Disk intensities from the Brault \& Neckel spectral atlas, Average SOT intensities measured on 2011 February 14, and the resulting SOT conversion factors}}}
\begin{tabular}{cccc}
\tableline
\tableline  
\multicolumn{1}{c}{Waveband} 	& Av. Disk Int.	& SOT Int.				&Conversion Factor  	\\ 
						& (W~cm$^{-2}$~sr$^{-1}$~\AA$^{-1}$)	&(DN~s$^{-1}$px$^{-1}$)	\\		
\tableline 
Red 						&0.2742		&36023.8				&7.6122$\times10^{-6}$	\\
Green 					&0.3541		&24236.6				&1.4610$\times10^{-5}$ 	\\
Blue 						&0.4316		&22558.1				&1.9133$\times10^{-5}$	\\
\ion{Ca}{2} H line			&0.0585		&2177.5				&2.6885$\times10^{-5}$	\\
\tableline
\end{tabular}
\label{table:conv_factors}
\end{center}
\end{table*}

\subsubsection{Optical Continuum}
\label{optical_continuum}

WL continuum sources (RGB) are difficult to detect against the bright photospheric background, and their identification required several steps of imaging processing, described in \cite{kerr14}, to be carried out. Newly brightened flare sources were identified in each frame (c.f. Figure 3 in \citealt{kerr14}) and Figure~\ref{sot_wl_image} shows them overlaid onto a continuum image, where the color refers to the time at which the sources were first identified as brightening.

\begin{figure}
\centering
\includegraphics[width = 0.5\textwidth]{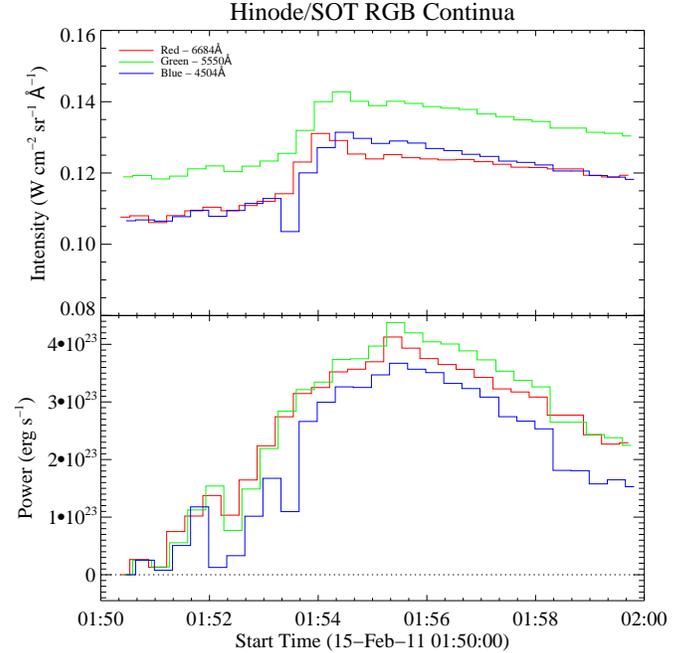}
\caption{Top panel: sample lightcurves of WL sources identified as first brightening near the peak of the flare. Bottom panel: background subtracted lightcurves. In both cases the units refer to the 4\AA\ passbands of the RGB filters.}
\label{sot_rgb_ltc}
\end{figure}

For every frame in which newly brightened WL sources were identified, a lightcurve for the RGB data was measured. The top panel of Figure~\ref{sot_rgb_ltc} shows sample lightcurves for newly brightened sources near the peak of the flare. As can be seen from the lightcurves, the sources had a quick rise to a peak intensity followed by a more gradual cooling period. The bottom panel shows the background subtracted summation of all identified brightenings converted to units of power in each 4\AA\ passband.

The instantaneous power, $P_\lambda$, emitted by these sources was calculated using $P_\lambda = \pi I_{f,\lambda} A \Delta \lambda$~\ergs, where $I_{f,\lambda}$ is the flaring intensity of the source in the SOT channel at wavelength $\lambda$ (that is, the background subtracted intensity of the source), $A$ is the area of flaring source in cm$^{-2}$, and $\Delta \lambda$ is the width of the passband of the SOT channel.
	
To find the total instantaneous power emitted as a function of time, the individual lightcurves from each of the identified sources were summed, as shown in the bottom panel of Figure~\ref{sot_rgb_ltc}. The peak power in each narrow 4\AA~passband was found to be on the order of $\sim4\times10^{23}$~\ergs. 

\subsubsection{\ion{Ca}{2} H Line}
\label{ca_ii_h}
The observations of the \ion{Ca}{2} H line show two clearly defined flare ribbons that had already brightened by the time that observations began (Figure~\ref{hsi_sot}). These ribbons evolve as the flare progresses, with the brightest part of the eastern ribbon moving northward, and that of the western ribbon moving southwards. The western ribbon appears brightest in two portions, with a region of lower intensity between the two. In the later stages of the observations there also appears to be some low intensity flows between the ribbons. 

\begin{figure}
\begin{center}
\includegraphics[width = 0.5\textwidth]{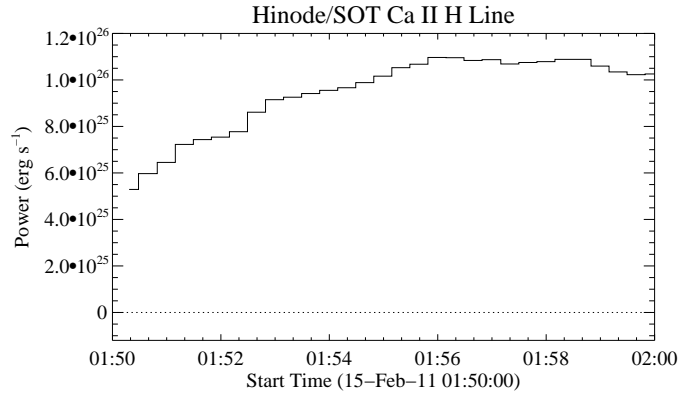}
\caption{Instantaneous power emitted from the \ion{Ca}{2} H line sources.}
\label{sot_caii_ltc}
\end{center}
\end{figure}

Flaring sources were identified by selecting pixels with an intensity greater than the mean-plus-10$\sigma$, where the mean and standard deviation, $\sigma$, were defined from a region of quiet Sun (the area between $\sim$160\arcsec\ and $\sim$240\arcsec\ in the Solar X-direction and between $\sim$-200\arcsec\ and $\sim$-180\arcsec\ in the Solar Y-direction).  

The total instantaneous power emitted by flare-enhanced sources was calculated by summing the power emitted by each flaring pixel using the same expression for $P_{\lambda}$ quoted in Section~\ref{optical_continuum}. Background intensity was measured by taking the mean intensity of the same quiet sun region defined above. The power emitted is shown in Figure~\ref{sot_caii_ltc}, and is generally on the order $\sim$10$^{26}$~erg s$^{-1}$. Integrating over time, the energy radiated as \ion{Ca}{2} H line emission was determined to be $\sim5\times10^{28}$~erg.

By the time that SOT began taking data in response to the flare trigger, the calcium ribbons were already flaring, which is why the background-subtracted power is significantly above zero at the beginning of the observations. The \ion{Ca}{2} H line intensity had not returned to background levels before the end of the SOT observations, with material continuing to radiate while cooling, so the value of energy quoted above is a lower limit on the total energy.

\section{Results}
\label{results}

\begin{figure*}[!t]
\begin{center}
\includegraphics[width=0.75\textwidth]{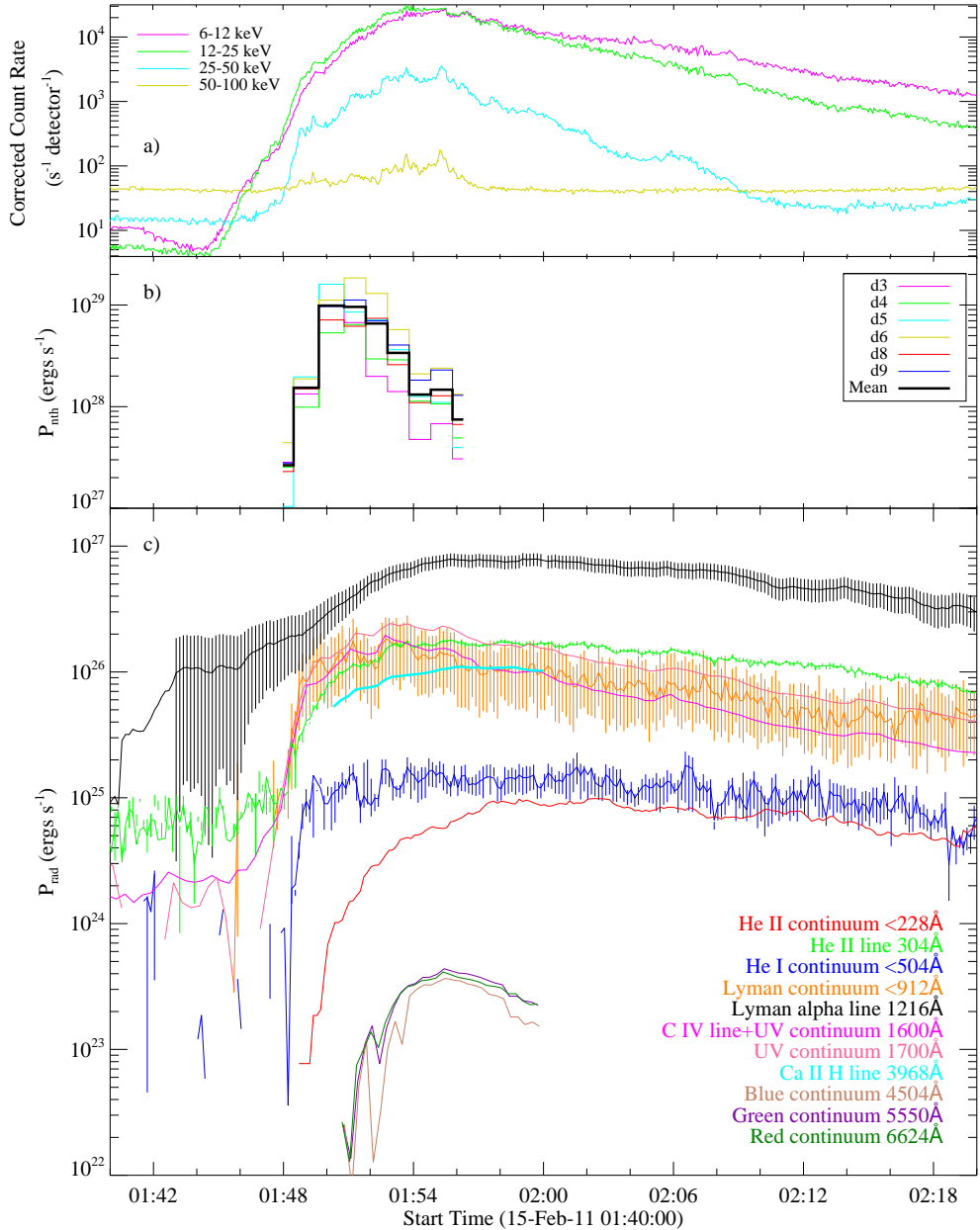}
\caption{a) Corrected count rates of HXRs from RHESSI in 4 energy bands: 6--12~keV (purple); 12--25~keV (green); 25--50~keV (cyan); 50--100~keV (yellow). b) power contained in nonthermal electrons (P$_{nth}$) compiled from individual RHESSI detectors (colored lines). The mean nonthermal power is shown in black. c) Time profiles of the radiated power (P$_{rad}$) of both line and continuum emission from the chromosphere in the optical, UV, and EUV. The \ion{He}{2} continuum lightcurve has been smoothed by 120~s for clarity.}
\label{flare_ltc}
\end{center}
\end{figure*}

In this section, the energetics derived from each emission process analysed individually in Section~\ref{data_anal} are combined and compared. Figure~\ref{flare_ltc}a shows the corrected count rate HXR time profiles as measured by RHESSI for the main phase of the event to guide the reader. The evolution of the power contained in the nonthermal electrons, as derived from RHESSI data using the methods described in Section~\ref{rhessi}, is shown in Figure~\ref{flare_ltc}b. Power derived from each individual detector (3--6, 8, and 9) is shown as a colored histogram, while the mean value at each time interval is plotted as a thick, solid black curve. This profile shows a rapid increase in electron power over the first 3 minutes of the flare, peaking at $\sim1\times10^{29}$~\ergs. By integrating under this curve, the total energy contained in the electron beam that was deposited in the chromosphere over the 10-minute impulsive phase was found to be $2\times10^{31}$~erg, which should be taken as a lower limit as the low-energy cutoff is an upper limit.

The bottom panel of Figure~\ref{flare_ltc} shows each of the lightcurves (also in units of \ergs) derived in Sections~\ref{eve}, \ref{aia}, and \ref{sot}. Almost all of the profiles peak around the time of maximum energy deposition. The \lya curve peaked $\sim$5~minutes later, while the \ion{He}{2} continuum had a more gradual rise phase due to the presence of a strong, underlying free-free continuum in the flare's early stages. Hinode/SOT was not observing at high cadence in its white light channels during the early impulsive phase and so data were only available from 01:50--02:00~UT. By summing the integrals under each lightcurve, the total energy radiated by the chromosphere as measured by EVE, AIA, and SOT was calculated to be $3\times10^{30}$~erg. This equates to about 15\% of the total energy contained in the nonthermal electrons as described above. 

\begin{table*}[!t]
\begin{center}
\small
\caption{\textsc{\small{Wavelength Range, Integration Time, and Total Energy Radiated for each of the Processes Presented}}}
\label{flare_energy_table}
\begin{tabular}{lccccc} 
\tableline
\tableline
\multicolumn{1}{c}{} 		&Wavelength Range (\AA)	&$\Delta\lambda$(\AA)	&Integration Time (UT)	&$\Delta$t (s)	&Total Energy Radiated (ergs)		\\ 
\tableline
\lya line				&1170--1270	 			&100					&01:40--02:30  			&3000		&$1.2\pm0.3\times10^{30}$		\\
He II line		    		&302.9--304.9  				&2					&01:44--04:00			&4560		&$3.4\pm0.1\times10^{29}$		\\
UV continuum			&1600--1740				&140					&01:44--03:00			&4560		&$2.6\times10^{29}$				\\
C IV line+UV continuum	&1464--1609				&145					&01:44--03:00			&4560		&$1.7\times10^{29}$				\\
Lyman continuum		&504--912         			&408					&01:44--03:00			&4560		&$1.8^{+1.0}_{-0.9}\times10^{29}$	\\
Ca II H line			&3967--3970				&3					&01:50--02:00			&600			&$5.5\times10^{28}$				\\
He I continuum			&370--504					&134					&01:44--03:00			&4560		&$3.0^{+0.6}_{-0.9}\times10^{28}$	\\
He II continuum			&200--228					&28					&01:50--02:40			&3000		&$1.6\times10^{28}$ 			\\
Green continuum		&5548--5552				&4					&01:50--02:00			&600			&$1.5\times10^{26}$				\\
Red continuum			&6682--6686				&4					&01:50--02:00			&600			&$1.4\times10^{26}$				\\
Blue continuum			&4502--4506				&4					&01:50--02:00			&600			&$1.2\times10^{26}$				\\ 
\tableline
\normalsize
\end{tabular}
\end{center}
\end{table*}

\begin{figure*}[!t]
\begin{center}
\includegraphics[width=0.48\textwidth]{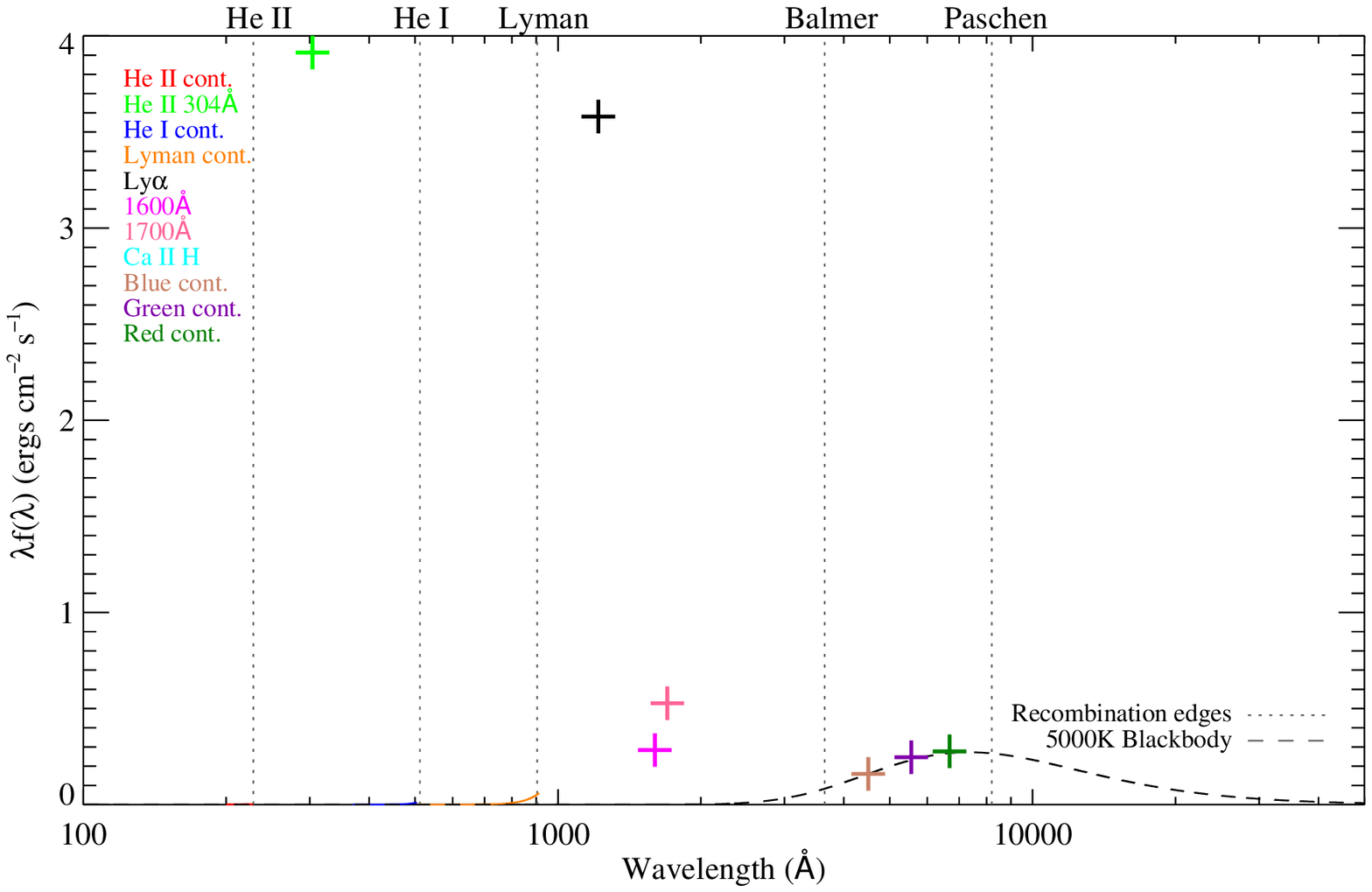}
\includegraphics[width=0.48\textwidth]{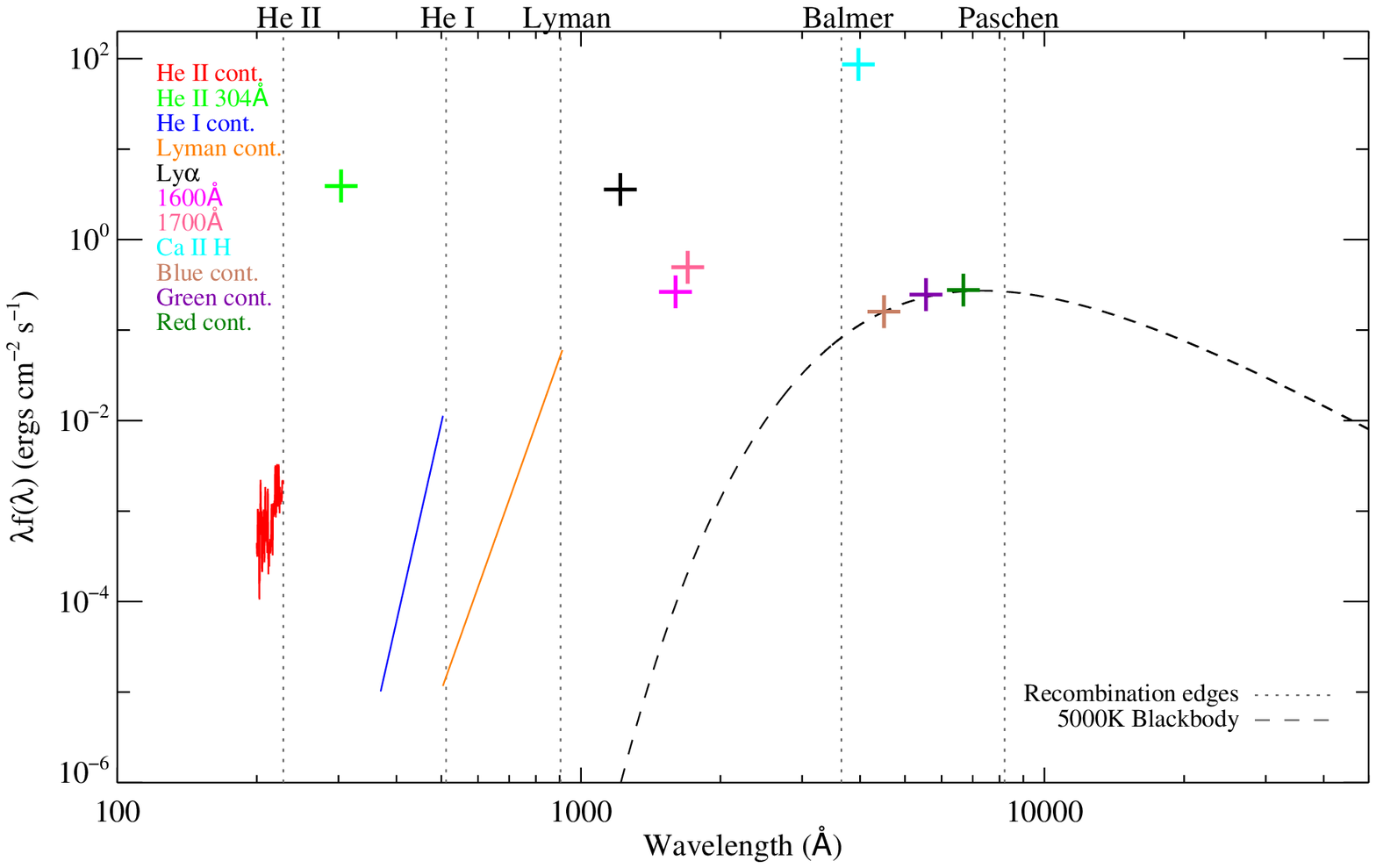}
\caption{Spectral Energy Distribution ($\lambda f(\lambda)$) of the flare excess energy plotted with both linear (left) and logarithmic (right) y-axes. The \ion{Ca}{2}~H line lies beyond the scaling of the left hand plot. The freebound continua from EVE are shown as solid lines while the other data are shown as colored crosses. Note that the symbols for the spectral lines and SOT points are larger than the actual bandpasses (see Table~\ref{flare_energy_table} for these values). The dashed black curve denotes a blackbody spectrum with a temperature of 5,000~K.}
\label{flare_energy_spec}
\end{center}
\end{figure*}

The \lya line was found to dominate the directly measured radiative losses, emitting $1.2\pm0.3\times10^{30}$~erg, which amounts to between 5-8\% of the total electron energy budget. The \ion{He}{2} 304\AA, the 1600\AA\ and 1700\AA\ channels, and the Lyman continuum were each found to radiate about 10$^{29}$~erg over the duration of the flare, while the \ion{Ca}{2} H line and the \ion{He}{1} and \ion{He}{2} continua each contributed $\sim10^{28}$~erg. The three narrow-band white light channels on SOT each contained $\sim10^{26}$~erg. Table~\ref{flare_energy_table} lists the bandwidths, integration times, and total energies of the different spectral samples. 

To visualize how the energetics derived from each instrument compare with one another, they are plotted on a spectral energy distribution (SED) plot ($\lambda f(\lambda)$). The SED plots of the flare excess energy for the 2011 February 15 flare are shown in Figure~\ref{flare_energy_spec} with both linear and logarithmic y-axes. The advantage of such a plot is that equal areas denote equal energies, for a linear Y-axis, so that the dominant contributions appear at a glance. Figure~\ref{flare_energy_spec} shows that a blackbody interpretation of the broad-band visible continuum might dominate, but that there are substantial gaps in the infrared, visible, and UV range that might turn out to be important.

\section{Summary and Conclusions}
\label{conc}

This paper presents a multi-wavelength analysis of the energy radiated by the lower solar atmosphere in response to energy deposited by nonthermal electrons during the impulsive phase of an X-class flare. The parameters of the nonthermal electrons assumed to be driving chromospheric heating and radiation were obtained using HXR data from RHESSI, while the corresponding response as determined from EUV, UV, and WL emission was measured by SDO/EVE, SDO/AIA, and Hinode/SOT, respectively. The total energy contained in the nonthermal electrons ($>2\times10^{31}$~erg) could comfortably account for that radiated away by the chromosphere over all passbands ($3\times10^{30}$~erg). By comparison, the amount of energy liberated by the eruption of this event has been calculated using Non-Linear Force Free Field extrapolations by \cite{sun12}, \cite{tzio13}, and \cite{asch14} who found that $3.7\times10^{31}$~erg, $8.4\times10^{31}$~erg, and $6.2\times10^{31}$~erg were released, respectively. Each of these values can satisfactorily account for the amount of energy that went into the accelerating the nonthermal particles.

While this study illustrates how energy deposited in the chromosphere during a major flaring event is re-radiated across the visible and E/UV parts of the solar spectrum, it says nothing about how the electron energy is transferred to the lower layers in particular, nor the depth at which the various line and continuum emissions originate. This can only be determined through imaging of near-limb events or advanced numerical modelling. The nonthermal electron parameters plotted in Figure~\ref{ebeam_params} and tabulated in Table~\ref{hsi_beam_params} therefore offer observational constraints with which to generate a chromospheric heating function based on actual measurements. The simulated solar atmosphere can then be directly compared with the observed lightcurves (and intensities) of chromospheric emission from SDO/EVE, SDO/AIA, and Hinode/SOT to reveal the physics of energy transport. There are, of course, several limitations to this approach. Current models are not yet capable of solving the equations of non-equilibrium and non-LTE optically-thick radiative transfer in multiple dimensions, and do not attempt to follow the structural changes (e.g., in MHD) of the flaring atmosphere. In reality, flares are complex and dynamic three-dimensional structures. Large X-class flares often comprise an arcade of loops, each heating a different part of the lower atmosphere. This is evident from footpoint motions such as those shown in Figure~\ref{sot_wl_image}. The observations presented here have been spatially integrated for ease of comparison. EVE does not provide any spatial information, and RHESSI does not resolve individual footpoints. Nonetheless, it would be informative to undertake such a comparison to see if any current models are capable of reproducing the observed lightcurves. Significant departures from observations may be attributed, first of all, to the uncertainty in the low-energy cutoff determined from RHESSI spectra. As mentioned in Section~\ref{rhessi}, this is an upper limit and therefore may be treated as a `free parameter' when used in modelling to bring the synthetic time profiles in-line with observations. Consequently, such an analysis could be seen as a consistency check on fits to RHESSI spectra: a match between theory and observations would imply that the measured value of $E_C$ was accurate. 

Another significant uncertainty lies in the area over which the energy is deposited. As can be seen in Figure~\ref{ebeam_params}e, areas derived from RHESSI and SOT images can differ by up to an order of magnitude. This can lead to similar discrepancies in the electron energy flux. If filling factors were revealed to be much less than unity in SOT images, then the energy fluxes could be in excess of 10$^{12}$~\ergcms, perhaps even closer to 10$^{13}$~\ergcms. If predicted and measured chromospheric emission still differ having exhausted some or all of these factors, then alternative energy transport mechanisms may need to be considered (e.g. Alfv\'{e}n waves; \citealt{flet08}). 

Multi-wavelength studies such as that presented here illustrate the value of bringing together observations over a broad spectral range, but also highlight the need for the inclusion of additional coverage, particularly those from ground-based instruments such as ROSA (\ion{Ca}{2}~K, \ha, blue continuum, G-band; \citealt{jess10}). Information on the Doppler velocity of the evaporating (or condensing) plasma from instruments such as the EUV Imaging Spectrometer \citep{culh07} onboard Hinode, or the recently launched Interface Region Imaging Spectrograph (IRIS; \citealt{depo14}) would give a more complete picture of the response of the lower solar atmosphere during solar flares. 

Despite this flare having great spectral coverage from space-based instrumentation, and at high cadence, only 15\% of the energy deposited in the lower solar atmosphere was observed. The inclusion of additional observations for future events would allow a more comprehensive SED plot to be compiled, and ideally, compared with bolometric measurements. This would help point to the `missing' 85\% of the energy and reveal how it is distributed throughout the solar atmosphere. Numerical modelling could also help in this regard; by accurately reproducing the observations that were available, they should be able to predict those that were not.

\acknowledgments
\noindent
This research was a result of several stimulating discussions between participants at a meeting on Chromospheric Flares held at the International Space Science Institute (ISSI) in Bern, Switzerland. ROM is grateful to the Leverhulme Trust for financial support from grant F/00203/X, and to NASA for LWS SDO Data Analysis grant NNX14AE07G. GSK acknowledges financial support of a postgraduate research scholarship from the College of Science and Engineering at the University of Glasgow. LF acknowledges financial support from STFC grant ST/l001808. The research leading to these results has received funding from the European Community's Seventh Framework Programme (FP7/2007-2013) under grant agreement no. 606862 (F-CHROMA).

\appendix
\section{RANSAC: RANdom Sample Consensus}

In order to reliably fit the SDO/EVE free-bound continua in the MEGS-B spectral range, the RANSAC (RANdom Sample Consensus; \citealt{fisc81})\footnote[3]{See also www.cse.yorku.ca/$\sim$kosta/CompVis\_Notes/ransac.pdf?} method was employed. This is an iterative method to estimate parameters of a mathematical model from a set of observed data that contains outliers. In the case of EVE data, emission lines that lie above the continuum were taken to be outliers, while the data points that comprise the continuum itself were considered as inliers. The procedure for implementing RANSAC is as follows:

\begin{figure}[!t]
\begin{center}
\includegraphics[width=0.60\textwidth]{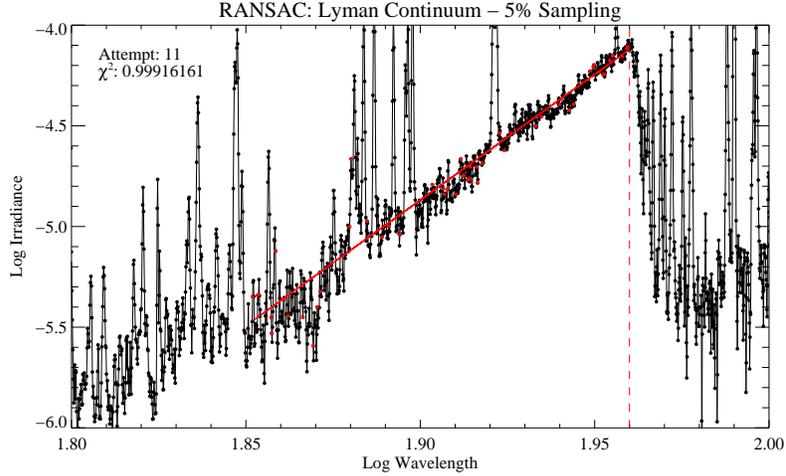}
\caption{Section of the EVE MEGS-B spectrum that contains the Lyman continuum plotted in log--log space. The red data points mark the 5\% selected at random to be fitted with a power law function (solid red line). In this example, an acceptable reduced $\chi^2$ ($<$1) was obtained on the 11$^{th}$ attempt. The vertical red dashed line marks the Lyman recombination edge at 912\AA.}
\label{ransac_random}
\end{center}
\end{figure}

Given a set of data points, $P$, randomly select a subset, $S$ (in this case $S=0.05P$; see Figure~\ref{ransac_random}) and fit with a chosen function (in this case a power law of the form $I=a\lambda^b$). Repeat $N$ times until an acceptable reduced $\chi^2$ value (i.e. $<1$) is reached. (In Figure~\ref{ransac_random}, N=11.) The number of attempts, $N$, is chosen to be sufficiently high to ensure that the probability, $p$, of at least one of the random samples does not include an outlier (typically, $p=0.99$). If $u$ is the probability that a given data point is an inlier, and $v$ the probability of any given data point being an outlier ($v=1-u$), then:

\begin{equation}
1-p=(1-u^{m})^{N}
\end{equation}
where $m$ is the number of data points in $S$. Therefore:

\begin{equation}
N=\frac{\mbox{log}(1-p)}{\mbox{log}(1-(1-v)^m)}
\end{equation}

\begin{figure}[!b]
\begin{center}
\includegraphics[width=0.60\textwidth]{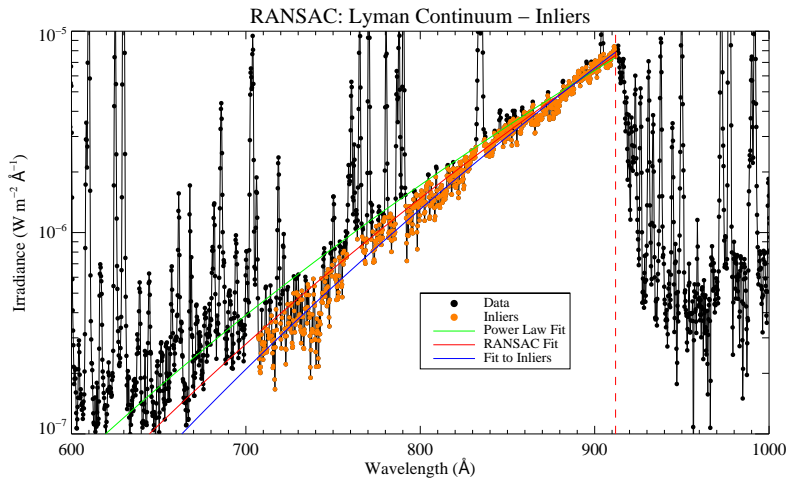}
\caption{The same section of the EVE MEGS-B spectrum as in Figure~\ref{ransac_random}. The orange data points mark the inliers: those which lie within 0.01 dex of the fit generated by the RANSAC method (red curve). The blue curve marks the fit to these inliers (and extended to shorter wavelengths), while the green curve is a straight forward fit to {\it all} the data points in the 700--912\AA\ wavelength range for comparison. The vertical red dashed line marks the Lyman recombination edge at 912\AA.}
\label{ransac_inliers}
\end{center}
\end{figure}

Having reached an acceptable $\chi^2$ after $N$ iterations, identify all the data points that lie within some predefined threshold, $d$, of the fit (for the Lyman continuum $d=0.01$~dex). These are the inliers. Fit inliers with chosen function. Figure~\ref{ransac_inliers} shows the same EVE MEGS-B spectrum shown in Figure~\ref{ransac_random} with the inliers highlighted in orange. The fit to these inliers, and extrapolated to shorter wavelengths, is overlaid in blue. For comparison, the fit to the randomly selected subset in Figure~\ref{ransac_random} is overlaid in red, while the green curve illustrates the fit to {\em all} the data points in the 700--912\AA\ wavelength range. The same process was applied to the \ion{He}{1} continuum blueward of the recombination edge at 504\AA, and to the upper and lower limits of both continua at each time step, as shown in Figure~\ref{eve_megsb_cont_fits}.

\end{document}